\documentclass[showpacs,preprintnumbers,prl,twocolumn]{revtex4}
\usepackage{amssymb}
\usepackage{amsmath}
\usepackage{graphicx}
\usepackage{bm}
\usepackage{amsfonts}

\begin{document}
\preprint{}

\title
{UCSB final report for the CSQ program: \\
Review of decoherence and materials physics for superconducting qubits }

\author{John M. Martinis}
\email{martinis@physics.ucsb.edu}
\author{A. Megrant}

\affiliation
{University of California Santa Barbara}%



\begin{abstract}
We review progress at UCSB on understanding the physics of decoherence in superconducting qubits. Although many decoherence mechanisms were studied and fixed in the last 5 years, the most important ones are two-level state defects in amorphous dielectrics, non-equilibrium quasiparticles generated from stray infrared light, and radiation to slotline modes.  With improved design, the performance of integrated circuit transmons using the Xmon design are now close to world record performance: these devices have the advantage of retaining coherence when scaled up to 9 qubits.

\end{abstract}

\volumeyear{year}
\volumenumber{number} \issuenumber{number}
\eid{identifier}
\date{\today}

\maketitle

\section{Introduction}

This is the final report for a five-year program at IARPA on coherent superconducting qubits (CSQ).  It summarizes our present understanding at UCSB of decoherence in a way that describes what is known about the underlying materials physics.  We hope this report will also inform the research community about current important issues, what research directions may be fruitful in the future, and even how to best design scalable qubit devices to build a quantum computer.

We would like to thank IARPA for the support of this research.  Their decision to improve materials has made a large impact on the field of superconducting qubits, to the point where we are now conducting experiments that are at the leading edge of quantum computer research.  In the past 5 years, we think the progress of superconducting qubits has exceeded that of other technologies because of the basic science learned in this program.

Decoherence progress for superconducting qubits is typically explained by listing the best coherence times $T_1$ or $T_2$ versus time, which has shown rapid and sustained improvement.  Although this represents the amazing progress of researchers in our field during the past 5 years, we want to emphasize that it is the understanding of the basic physics behind these improvements that are the key to future improvements, mostly because the next research frontier is knowing how to scale: Building and optimizing complex qubit systems will certainly require various trade-offs that can only be optimized by full knowledge of the underlying physics of coherence.  For example at UCSB, such data has been foundational to the rapid development of the Xmon transmon \cite{xmon} during the last 1 1/2 years, where we have progressed from 1 qubit to now 9 qubits with no apparent degradation of coherence.

We will also explain how plots of coherence versus time can be a bit misleading, because of important issues like scaling and parameter choice.  The idea here is that summarizing the performance of a complex system like a qubit is almost impossible with a single number; this review hopes to dive into some of the subtle issues behind a full understanding of coherence.

Please note this is a report of UCSB research over the last 5 years, not a review of the entire field.  We will comment on other work as appropriate to summarize what we understand about the basic materials physics.  We welcome comments and debate about what is communicated here, and can revise this document to make it better.

This review will be organized by sources of decoherence.  This may be readily categorized by considering a superconducting qubit as a nonlinear LC resonator, where loss comes from the capacitor, the inductor (including the Josephson inductance), and radiative loss from the embedded circuit.

\section{capacitor loss}

The importance of dielectric loss in capacitors was understood at the beginning of the CSQ program.  Although new materials were investigated and there is a deeper  understanding of its physics, all of the advances from capacitor loss in the last 5 years came by avoiding lossy dielectrics as much as possible.  This concept has been well expressed by the Yale group by considering the participation ratio of dielectric materials.  Lowering their participation was most notably achieved with their invention of the 3D transmon, where the participation ratio was made as high as possible for the surrounding vacuum because it has no dielectric loss.

As dielectric materials are always present, for example in the substrate, it is vitally important to measure their loss.  To our knowledge, dielectric loss at low temperature arises from the presence of two-level states (TLS) formed by random bonds that tunnel between two sites.  Table\,\ref{tab:losstangent} shows what is presently known at UCSB for both amorphous and crystalline materials; the intrinsic loss tangent data is taken at low temperature and power where the TLS are not saturated, as appropriate for qubit experiments.

\begin{table}[t]
\caption{\label{tab:losstangent} Table of intrinsic dielectric loss tangent $\tan\delta_i$ for amorphous and crystalline materials.  The limit of the intrinsic loss of bulk Si has not been measured well; its limit comes from coplanar resonators recently made at UCSB \cite{megrant14}. }
\begin{tabular}{c|c|c}
\hline
\hline
amorphous & reference &  $\tan\delta_i \times 10^6$ \tabularnewline \hline
MgO & \cite{oconnell} & 5000-8000 \tabularnewline
PECVD SiO$_2$ & \cite{oconnell} & 2700-2900 \tabularnewline
AlN & \cite{oconnell} & 1100-1800 \tabularnewline
AlO$_x$ & \cite{martinis}  & 1600 \tabularnewline
ebeam resist & \cite{quintana}  & 510-770 \tabularnewline
sputtered Si & \cite{oconnell} & 500-600 \tabularnewline
thermal SiO$_2$ & \cite{oconnell} & 300-330 \tabularnewline
spin-on teflon& \cite{mcdermott} & 140-180 \tabularnewline
SiN$_x$ & \cite{oconnell} & 100-200 \tabularnewline
a-Si:H  & \cite{oconnell} & 22-25 \tabularnewline
teflon & \cite{evangeller} & 1-2 ? \tabularnewline \hline
crystalline \tabularnewline \hline
YSZ & \cite{yiyin} & 400-500 \tabularnewline
LSAT & \cite{yiyin} & 70-90 \tabularnewline
MAO & \cite{yiyin} & 12-18 \tabularnewline
SLAO & \cite{yiyin} & 10-12 \tabularnewline
YAG & \cite{yiyin} & $< 10$ \tabularnewline
YAO & \cite{yiyin} & $< 10$ \tabularnewline
LAO & \cite{yiyin} & $< 10$ \tabularnewline
Si & \cite{megrant14} & $<0.15$ \tabularnewline
sapphire & \cite{creedon} & 0.02 \tabularnewline
\hline
\hline
\end{tabular}
\end{table}

Note that the intrinsic loss tangent of crystalline Si, using ``undoped'' material with resistivity $> 1000\,\Omega-\textrm{cm}$, has not yet been measured well for bulk Silicon.  It is known to be a low-loss substrate since it produces coplanar resonators with loss a factor of $\sim 2$ better than for a sapphire substrate \cite{megrant14}.

The other interesting material is teflon, an amorphous material with extremely low loss, which was measured in a preliminary manner using a half-wave resonator made from commercial Nb semirigid coax.  The low loss is presumably arising from the absence of hydrogen in the material; the random bonding of fluorine to carbon likely does not produce TLS in the GHz range because the fluorine is much heavier than hydrogen, inhibiting tunneling.  Teflon was explored a bit in this program, but because it does not adhere well to substrates and is soft, it is not considered a practical material for our current process.

One concern is the need to purchase low-loss Si or sapphire substrates.  To our knowledge there is no publicly known test data or procedure for specifying the quality of substrates or for easily measuring the loss when received from vendors.  Our current plan is to purchase Si wafers by the boule (100-200 wafers) and then test the quality of the batch by fabricating coplanar resonators on one wafer.

At the start of this research program, we investigated making parallel plate capacitors with low dielectric loss.  We had to abandon this approach at the end of the program as it seemed too hard.  Parallel plate capacitors are still an interesting line for future research, but we expect it probably will only be restarted once the field matures to where these structures are desperately needed.  Moving our research from the phase qubit to the transmon qubit meant that we did not need large capacitors.  This implied that we could no longer use direct coupling over long distances, but this choice was consistent with the surface code architecture where only nearest neighbor interactions are required \cite{fowler}.

Transmons only need modest capacitance that can easily be made using coplanar (interdigitated) capacitor structures, with the capacitance mostly coming from the low loss substrate.  A key part of our program was investigating the properties of these capacitors using coplanar resonators, which only requires one layer of fabrication.  In the past 5 years we have measured many hundreds of devices: Fast fabrication and turnaround, giving detailed knowledge based on actual loss measurements, were key to understanding the materials physics.

In this program we also understood a better way to analyze resonator data that allows a direct measurement of the internal loss (quality factor $Q_i$) without having to measure and subtract the coupling $Q_c$ \cite{megrant}.  After testing hundreds of devices using this methodology, we have found it gives consistent results for ratios $Q_i/Q_c$ as high as 10, although we recommend this ratio be no larger than about 2 for maximum reliability.

With low loss substrates, the loss in coplanar capacitors is dominated by amorphous dielectrics at interfaces, coming from the vacuum-metal, metal-substrate, or vacuum-metal surfaces \cite{wenner}.  The vacuum-metal interface can be ignored since it is a minor contributor, based on the mismatch of dielectric constants.  For the remaining two interfaces, this model gives roughly equal loss contribution assuming similar interface thicknesses and dielectric constants of 10. This interface model has total loss that scales inversely with system size, which has been verified with resonators made at UCSB with different characteristic gaps and center-line widths.  Note that the resonator quality factor and qubit $T_1$ may start to saturate around gap widths of $50\,\mu\textrm{m}$, presumably because other loss mechanisms like radiation begin to dominate.

Because the metal-substrate interface is buried, the loss from this interface can be minimized through careful fabrication techniques \cite{megrant}.  We found that aggressive ion-milling of the substrate before Al deposition produces an amorphous layer at the interface, increasing loss by about a factor of 2 \cite{quintana}.  Using our MBE system, we found that cleaning the sapphire substrate with a high temperature anneal gave lower resonator loss.  Annealing in an plasma/atomic oxygen source also further reduced loss, presumably by better cleaning or not allowing oxygen to diffuse out from the surface, amorphizing the sapphire.  We believe it may possible to anneal the wafers simply in O$_2$ to produce similarly good results, based on a series of \textit{in situ} XPS measurements.  This interface cleaning gave an improvement in resonator Q of about a factor of 2 with respect to simple evaporation of Al in a $10^{-7}$ Torr background, i.e. in our junction evaporator system.  This improvement with MBE Al is consistent with removing most or all of the loss from the metal-substrate interface.  We think it is possible to lower the loss from the remaining vacuum-surface interface with more advanced fabrication ideas, so this is an interesting area of future research we are working on now.

We note that patterning the Al layers via liftoff, common in many qubit groups, likely produces additional loss due to ebeam resist residue \cite{quintana}.  We find that oxygen descum processing before the evaporation step can remove this layer for significant improvements in loss.  In this work we also show that these contaminant layers may be detected and measured using ellipsometry techniques.

These interface layers give loss in qubits, and our present limits of $T_1$ are consistent within a factor of two of loss measured in resonators \cite{xmon}.  We additionally find that $T_1$ varies by about a factor of 2 to 4 with qubit frequency, a measurement that cannot be done in resonators because they have fixed frequency.  We find this change is consistent with a model of TLS defects on the interfaces; far away from the edges the large density but small coupling gives a background $T_1$ decay, but the few states near the edges that are well coupled give peaked $T_1$ decay as the qubit moves into and out of TLS resonance.

We note that other groups doing transmon research have not reported significant effects from TLS.  This is likely because most other groups are not fabricating tunable transmon devices for the latest generation of long $T_1$ devices, so that variations in $T_1$ are not directly measureable.  However, when looking at $T_1$ data of these other devices, one finds that a range of $T_1$ is reported, which seems compatible with our spread.  We thus surmise that all transmons are susceptible to variations in $T_1$, either observing it through a device change, or for us, a frequency change.  Clearly, tunable qubits allow this effect to be explored and understood in much greater detail.

We observe that the $T_1$ variation with frequency, the ``$T_1$ spectrum'', changes with time.  The fluctuations of the presence of TLS defects is conventionally understood as each TLS being coupled to nearby TLS through the crystal strain field, which effectively turns on and off individual fluctuators.  We often fine tune the qubits on a daily basis to get best behavior.

This behavior of fluctuating TLS \cite{neill} was studied in this program in great detail by looking at the changes with time of a resonator frequency and its Q.  These fluctuations have long been studied for Microwave Kinetic Inductance Detectors (MKID) devices, but here we were able to connect the fluctuation behavior with a more detailed microscopic model that was consistent with TLS loss from the interfaces.  We find that the parameter regime of current resonators and Xmon qubits are near the ``statistically avoided'' limit, so small improvements in interface quality might yield large improvements in device performance.

At the beginning of the program we understood that TLS loss in the capacitor of a large area Josephson tunnel junction significantly damps the qubit \cite{martinis}.  For large junctions with area greater than about $10\,\mu\textrm{m}^2$, they behave like a normal resistive loss tangent with $\tan\delta_i \simeq 1.6\times10^{-3}$.  For small junctions less than about $1\,\mu\textrm{m}^2$, the number of defects are few enough so that they are mostly statistically avoided, giving the capacitor no loss. The simple TLS model is consistent with all data we have seen in the past 5 years.

During the program, we also investigate the statistical distribution of coherence times of TLS in the junctions, which is consistent with simple models for phonon radiation \cite{shalibo}.

In an early program review, we mentioned that one expected to occasionally see a TLS in transmon devices.  Researchers from Yale then corrected me by responding that they did not see TLS effects.  In discussion afterwards, it was hypothesized that the small junction area maybe allows the junction to relieve the stress in some way, effectively annealing the TLS defects.  During subsequent work with transmons at UCSB, we have been looking for large splittings in spectroscopy that would indicate the presence of TLS defects in the junctions.  By making many qubit devices and performing $T_1$ spectroscopy over wide frequency ranges, we did find several candidates for TLS in the junctions.  However, their density seems to be lower than predicted by simple scaling of the junction area; our present thought is that they are found at about 1/3 to 1/8 the density expected from prior models made from phase qubits, but they are still there.  This clearly is an important issue for scaling up to a large quantum computer, and in future research the statistics need to be better quantified and understood.

Although we are no longer incorporating a-Si:H in qubits, this material and its multilevel process are still being used to fabricate low loss dielectrics for parametric amplifiers.  Our first device used multilevel lithography to fabricate an on-chip capacitor with a single ended (not differential) signal line, with a separate input for flux-pumping \cite{mutus13}. This design simplified operation of the Josephson parametric amplifier and brought better performance because the input impedance was reduced by a factor of 2.  Our next generation paramp used an impedance transformer, made from a tapered transmission line, which gave much higher bandwidth and saturation power \cite{mutus14}. The ability to reliably use multilayer metallization with complex layout was key to improvements of UCSB paramps.

Currently we are building and testing a travelling wave parametric amplifier, in collaboration with J. Gao at NIST, Boulder.  The use of the low-loss dielectric a-Si:H is a key technological improvement here, since prior work at Berkeley and Lincoln labs has shown poor performance because of high loss from their SiO$_x$ insulators.  For a-Si:H we observed in transmission measurements negligible microwave loss, less than about 0.2 dB, limited seemingly by connector reflections, and find proper functioning of the amplifier with near quantum limited noise \cite{white}.  This device has many thousands of junctions and capacitors, and thus severely tests the reliability of all process steps.  We have found the a-Si:H has a small probability to form cracks at edges, so we have optimized our design to reduce the number of edges.  Note that stress and crack formation may be a practical limitation to low-loss dielectrics, since it is the increasing coordination number of SiO$_x$ to SiN$_x$ to a-Si:H that is thought to lower loss, but the larger number of bonds overconstrains the atomic positions, creating larger internal stress.

\section{Inductor and Junction Loss}

In the early stages of this program, we found that resonator $Q$ measurements were unreliable once we started looking at devices at the $Q=300,000$ level.  Our measurements became reliable by removing loss from trapped vortices and quasiparticles, which then enable a series of detailed experiments understanding dielectric loss.  We first discuss here our latest understanding of these two issues.

Trapped vortices in superconducting films provide a site for loss due to the normal core of the vortex and its motion \cite{song}.  For a film of width $w$, vortices will be trapped in the film when cooled through the superconducting critical temperature with magnetic fields $B_c \gtrsim \Phi_0/w^2$  \cite{stan}.  When vortices form in the center line of a coplanar resonators, the loss is large, whereas vortices in the ground plane are less coupled.  For example with centerline widths of $30\,\mu\textrm{m}$, the high power quality factor is $Q_{HP} = 14\,\times 10^6$ at $B_c<1\,\textrm{mG}$ but decreases slightly to $Q_{HP} = 3.8\,\times 10^6$ at $B_c=7\,\textrm{mG}$, whereas loss increases rapidly above 50\,mG when vortices begin to enter the center line.

To circumvent these problems, we incorporate mu-metal shields around our dilution refrigerator and device mounts to reduce the magnetic field to less than about 1 mG.  Additionally, we use non-magnetic screws and SMA connectors for parts inside the shield, and have a dedicated test setup for screening.

It is possible to relax the requirements for magnetic shielding by placing holes in the ground plane.  The stray fields are then trapped in the hole, eliminating the normal core and its dissipation.  This solution has worried us during the last 2 years, as it is possible that the additional edges of the hole introduce a surface where there is TLS loss.  We have recently tested this hypothesis, and find that TLS loss does not increase with the use of holes in the groundplane \cite{chiaro}.  We now feel confident that holes are an acceptable solution, and by doing so the resonators become insensitive to stray fields up to about the 50\,mG when vortices start to form in the centerline.

In the last 5 years, the UCSB and Yale groups have performed much research on understanding dissipation from quasiparticles.  Theory has predicted that quasiparticle dissipation should vanish for a junction phase difference of $\pi$, which has been beautifully confirmed with a fluxonium experiment.  Experiments at UCSB have probed the increased dissipation and frequency shift with increased quasiparticle number \cite{lenander}, showing excellent agreement with theory.  We have also shown that non-equilibrium quasiparticles can excite qubits above their normal thermodynamic value \cite{wenner13}.

Although this is important fundamental physics, our main interest was practical: how to discover sources of non-equilibrium quasiparticles and understand ways to reduce it.  This was investigated with resonator samples since quasiparticles also reduce their quality factor.  We found that an overlooked source of energy that could break Cooper pairs was infrared radiation, which was probably not appreciated because it is very difficult to make microwave-tight seals at the appropriate infrared frequencies of 80 GHz.  In our paper, we also showed how infrared shielding could be tested by varying the temperature of a blackbody radiator at 4\,K, and tried a variety of shielding methods to discover what were the best designs \cite{barends}. A critical concept was to use multiple layers of shielding and absorbers, not relying on one good joint.  We put limits on the quasiparticle density from infrared sources.  We have also been incorporating infrared shielding, as low pass filters, on qubit lines.

We note that quasiparticle densities are still thought to be far above their equilibrium value, zero for nominal conditions.  There are many other sources of radiation that can break Cooper pairs, such as the slow (logarithmic) decay of energy with time that is well known in low temperature systems, presumably from the release of mechanical stress.

There was concern by some members of the superconducting qubit community that transmons should be isolated from the ground plane, in a non-galvanically coupled manner presumably to break diffusion.  With good performance of the Xmon, this idea does not seem to be important.

Concerning critical current fluctuations, we found the data from the University of Illinois to be very interesting and useful.  We have not seen any effects of dephasing from these fluctuations, but will continue to look for them.

Although most all of our recent research has focused on the use of Al metal, we did investigate the alloy TiN as a superconductor.  We found TiN to have acceptable low loss in resonators as we obtained a high $Q > 1\,\times10^6$ \cite{ohya}.  Unfortunately, we also found that TiN was hard to fabricate, as it seems to incorporate oxygen after venting from the growth chamber, producing significant variations in resistivity.  We note that Pappas at NIST is able to grow much more stable and reproducible films.  TiN is an interesting material since its high resistivity give high kinetic inductance, allowing for small resonator structures.  But since a key requirement now for multiplexed readout is the ability to reproduce resonant frequencies to within about 1 MHz, we do not believe this will be easy to do with TiN because of its sensitivity to resistivity.

Flux noise is an important dephasing mechanism in superconducting qubits.  Because of research at McDermott's laboratory at the University of Wisconsin, we now understand that the fundamental mechanisms behind the noise is surface spins.  The microscopic source of these spin fluctuations are now an active area of research, but recent surface-science suggest that absorbed O$_2$ or OH may be the source.  The reported magnitude of flux noise has decreased over the last 5 years, presumably due to improvements in fabrication quality.  At UCSB, we have pioneered the development of measurement techniques to determine the flux noise spectral density below 1 Hz using the qubits themselves \cite{sank}.

Many groups now only use fixed frequency transmons so that the dephasing coherence time $T_2$ is not degraded by flux noise.  The UCSB group uses transmons that are tunable since it allows the qubit to quickly change frequency, which rapidly turns on and off the coupling between neighboring qubits.  We observed shorter dephasing times because of flux-noise induced frequency noise, but this dephasing does not seem to be a serious problem for gate fidelity when using the surface code.  This can be understood by noting that the phase error for flux noise is proportional to time, and with qubit error proportional to the square of phase error, then the qubit error scales as the square of time.  With 1/e decay times due to flux noise in the 3-7\,$\mu$s range, the qubit error at gate times 10-40\,ns is negligible.  This physics is contrary to conventional wisdom and the CSQ program goals, where there is a concerted effort to maximize $T_2$ and use it as a fundamental figure of merit.  A full discussion of this idea, along with an experiment that precisely measures small qubit error at the appropriate small gate times, will be submitted soon \cite{omalley}.  In this recent work we find that the dominant dephasing mechanism is not 1/f flux noise, but a newly discovered two-level state type of defect that is not yet understood well.

\section{Radiation and wiring loss}

Qubit decoherence may also come from dissipation and noise from leads connected to the device.  Although the basic physics has been understood for some time, we have developed new ways to think about and calculate these effects, which we review here.

The effect of the environment on the qubit has often been described as the ``Purcell effect''.  While this phrase may nicely trace back to fundamental ideas by Purcell, we believe it is worthwhile to also advertise the paper by Esteve \textit{et. al.} published in 1986 as a more foundational theory for the field of quantum circuits \cite{esteve}.  The reference to the Purcell effect does not explicitly tell you how to calculate things, but simply relates the idea that decay rate is proportional to the density of states, given by the real part of the admittance function; this is obviously correct based on Fermi golden rule arguments.  In contrast, the Esteve article teaches how to calculate the effect of dissipation and dispersion on quantum circuits, in perturbation theory, as a function of the classical response of the environment.  Although motivated by the phase qubit, its results are expressed in terms of eigenstate transition frequencies and matrix elements, so it can be used for an arbitrary quantum circuit.  It also intuitively explains how for a linear circuit the standard results for electrical circuits are obtained, justifying the use of linear concepts for a weakly non-linear device such as the transmon.

It has also become common to use a double sided noise spectrum to represent dissipation, instead of an impedance or admittance function.  Of course this is correct because of the fluctuation-dissipation theorem, and many times a noise description makes theory more compact or easy to explain.  However, we have found discussions of experimental data with quantum noise spectral density to be much more confusing and often ill explained, since we have a much more intuitive understanding of the magnitude of dissipation if expressed in ohms.

\begin{figure}[t]
\includegraphics[width=0.45\textwidth]{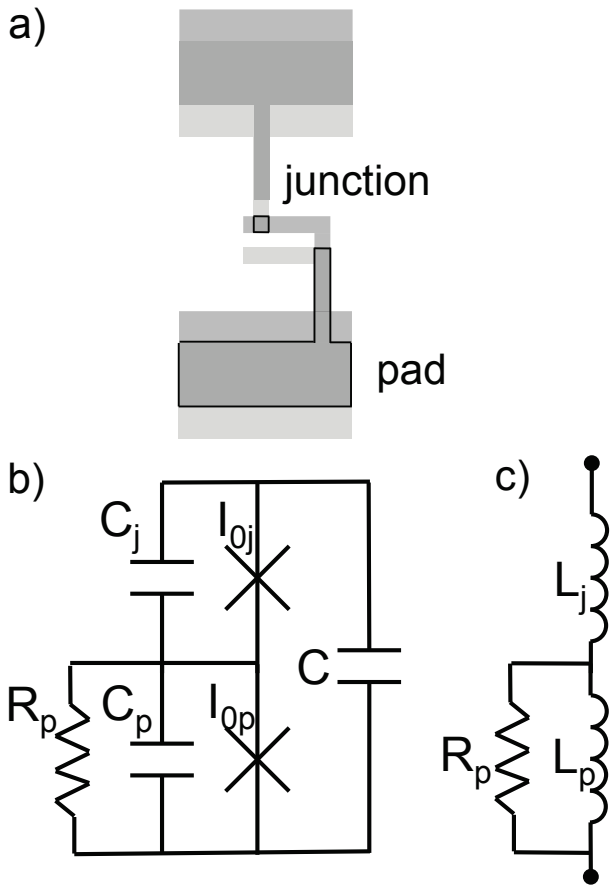}
\caption{\label{fig:transmon} (a) Physical layout of transmon junctions. When making a Josephson junction with double angle evaporation, the first Al layer makes good contact to the leads because of an ion-mill cleaning step.  The 2nd counterelectrode makes contact to the leads through a second large-area junction, labeled as pad. (b) Schematic of double junction circuit.  Josephson junction and capacitance for both the transmon and pad junctions are shown, along with dissipation $R_p$ coming from loss tangent of large area pad junction. (c) Effective linearized circuit for the double junction, which makes up the admittance $Y$.  $L_j$ is the transmon inductance whereas $L_p$ is the (smaller) pad inductance.
}
\end{figure}

To illustrate these ideas and discuss a more intuitive understanding of dissipation, we consider a transmon circuit with a stray junction.  As illustrated in Fig.\,\ref{fig:transmon}(a), the fabrication of the Josephson junction introduces a second series junction to the circuit, which acts as a contact pad having much larger area and critical current.  Although the inductance from this pad junction is small and thus has negligible effect on the qubit frequency, we are interested here in calculating dissipation effects.  Because the self-resonance frequency of each junction and its capacitor is much larger than the transmon frequency, the capacitance of each junction may be ignored.   As shown in Fig.\,\ref{fig:transmon}(b), we ignore the dissipation of the TLS defects in the capacitor for the transmon junction because it has small area.  But this can not be done for the large area junction; here, its large size implies that a simple resistor model $1/R_p= \omega C_p\tan\delta_i$ is a decent approximation.  Because the transmon is weakly non-linear, we can linearize the circuit response as in Fig.\,\ref{fig:transmon}(c) to then solve for the transmon decay rate.

\textbf{(1) Algebra.}  Qubit dissipation for this circuit can be solved using an algebraic calculation following the Esteve article \cite{esteve}. Using circuit parameters defined in the figure and the symbol $\parallel$ to represent the standard parallel computation of impedances, one finds the total admittance to be
\begin{align}
Y &= 1/[i\omega L_j + (i\omega L_p \parallel R_p)] \\
  &= \frac{1}{i\omega L_j + \frac{i\omega L_p R_p}{i\omega L_p + R_p}} \\
  &\simeq \frac{1}{i\omega(L_j+L_p)} + \frac{L_p^2}{(L_j+L_p)^2}\frac{1}{R_p} + ...
\end{align}
where the final equation is given in the small dissipation (large $R_p$) limit.  The first term is the total inductive response of both junctions, whereas the second describes the dissipation coming from the pad resistor.  Defining $1/R=\textrm{Re}\{Y\}$, the energy decay time is found  to be given by the harmonic oscillator formula \cite{esteve}
\begin{align}
T_1 &= RC \\
&\simeq (L_j/L_p)^2 R_p C
\end{align}
where we have assumed $L_j \gg L_p$.  Although conceptually this calculation method is simple, the algebraic manipulations for Y is quite tedious and gives one little intuition on how this circuit works.  We next explore other ways to look at this calculation to increasingly give more understanding of the physics.

\textbf{(2) Impedance transformation.}  This calculation can be understood more intuitively with the concept of impedance transformation between parallel and series circuits.  For an inductor or capacitor element with imaginary impedance $X$, in the small dissipation limit the quality factor of a parallel resistor $Q=R_p/|X|$ must be the same as for a series circuit $Q=|X|/R_s$.  Equating Q's give an impedance transformation $R_s=|X|^2/R_p$.  For our transmon circuit, the pad resistance is transformed into a series resistance $R_s=(\omega L_p)^2 /R_p$.  It can then transformed back into a parallel resistance across the entire transmon circuit
\begin{align}
R=(\omega L_j)^2/R_s = (L_j/L_p)^2 R_p
\end{align}
where we have again used $L_j \gg L_p$.  Here the pad dissipation is transformed up in impedance by the ratio $(L_j/L_p)^2$, leading to the above formula for $T_1$.

\textbf{(3) Thevenin-Norton equivalents.} The idea of this calculation is that dissipation can be represented by a quantum noise source with spectral density $S_I \propto 1/R_p$. Representing $R_p$ with a parallel noise source $I_n$, it can be replaced by an equivalent series (Thevenin) noise source has magnitude $(\omega L_p)I_n$, which then is equivalent to an equivalent parallel (Norton) noise source across both inductors, with magnitude $(L_p/L_j)I_n$.  Since the square of the noise is inversely proportional to the resistance, the reduction in magnitude of the noise is equivalent to an impedance transformation as described above in section 2.

\textbf{(4) Current divider.} The previous idea of impedance transformation from noise can be further simplified by calculating the noise current flowing through a short across the output $Y$ \cite{neeley}.  Here, the two junction inductances produce a divider so that only the ratio $L_p/L_j$ of the original noise flows through the shunt. This gives a noise ratio and impedance transformation as found in section 3.

\textbf{(5) Power dissipation} In this calculation, we consider that the transmon oscillates with voltage magnitude $V$ so that the energy stored is $CV^2$.  Because the two inductors form a voltage divider, the voltage across the pad inductor is $(L_p/L_j)V$, giving a power dissipation $P=(L_p/L_j)^2 V^2/R_p$.  The energy decay time is then computed using $T_1=E/P$, giving the above result.

We have found with Xmon transmons that proper design of the ground plane is critically important.  The problem is that chip wiring breaks the global connectivity of the ground plane, creating slotline modes that can be easily coupled to, so as to provide additional radiating modes.  In the Xmon paper we found that qubit coherence increased with increasing size of of the Xmon capacitor, up to a size when holes in the $T_1$ spectrum appeared, presumably coming from increased capacitance coupling to slotline modes \cite{barends}.  In the next generation of Xmon devices where we introduced crossover wiring between the ground planes, these modes were largely gone \cite{barends14}; The reduction of modes is clearly visible when comparing Fig.\,4 of \cite{barends} with Fig.\,S1 of \cite{barends14}.

In detail, we observed that the larger qubits had higher coherence times, but we also noticed some wideband modes $>10\,$MHz with decreased $T_1$ that persisted between cooldowns, suggesting they were not from TLS. Additionally, we saw the presence of modes where the equilibrium population of the excited qubit state would be greater than 30\%.
In subsequent cooldowns, we added a large number of wirebonds across breaks in the groundplane, which did not dramatically effect $T_1$ performance and did nothing for the hot modes.  But adding a 500 MHz low-pass filter on the Z line completely removed the hot modes.  As we further understood these experiments, we found two design errors: slotline modes and stray coupling in the bias line.  We then redesigned for the next chip adding SiO$_2$ crossovers and using a more symmetrically designed bias line that gave stray coupling with a $T_1 > 10\,m\textrm{s}$.  This changes gave the improvements reported in \cite{barends14}.

The solution to slotline modes has traditionally been the use of wirebonds.  As these wires are long and their impedance is roughly 20-30\,ohms at 6\,GHz, these are not a very good shorts for the ground plane.  Although they help performance, they are not an effective and well-engineered solution to this problem.  We have analyzed the performance of wire bonds and developed a better solution based on suspended air bridges \cite{chen}.  We used ion-milling to make low-resistance connections with the bridge wiring, and the added loss was measured with a controlled series of resonator experiments.  The fabrication process is somewhat delicate because the ion-milling hardens the resist making it hard to strip, but we find good fidelity and low loss when done properly.

Note that we initially fabricated crossovers on a sapphire wafer using liftoff Al over evaporated and liftoff SiO$_2$.  Because amorphous SiO$_2$ has a large density of charge defects that may polarize Si and give surface conduction, we had to develop airbridge technology for Si substrates.

When designing transmons and understanding the effects of slotline modes, it was important to understand capacitance coupling across the device.  Traditionally, numerical solvers have been used for this calculation and are known to work well, but we found this methodology gave us little design intuition especially as our circuits became more complex.  In thinking about the physics of this problem, we discovered a simple formula for capacitance that has proven to be very accurate and scalable to complex circuits \cite{martinis14}.  For a continuous ground plane interrupted by thin cuts to define electrodes, the capacitance from electrode 1 to electrode 2 is given by the simple integral
\begin{align}
C_{12} &= (\epsilon/\pi) \int \int dA_1\, dA_2\, /|r_1-r_2|^3 \\
&\simeq  (\epsilon/\pi) A_1 A_2/r^3,
\end{align}
where in the last formula we assume areas $A_1$ and $A_2$ are well separated and have an average distance between them given by $r$.  Here $\epsilon$ is the average dielectric constant for the substrate and vacuum.  We have found this formula to be extremely powerful for understanding and fixing problems of stray coupling in our resonator circuits, which has shown coupling Q's in excess of 3 to 10 million.

This idea has also been useful in understanding the coupling of the Xmon to box modes.  Here, the total capacitance of the island to the box mode can be calculated using a simple parallel plate formula $C_b = \epsilon A/t$, where $A$ is the island area and $t$ is the distance to the box lid.  This formula shows that one can decrease coupling to a box mode by several orders of magnitude by floating the substrate from the ground of the box, which both increases the distance and decreases the dielectric constant.  We note that this floating design had been previously understood as an optimal way to decrease crosstalk between microwave lines connected to the chip \cite{wenner11}.

Crosstalk between qubits has been measured for a 5 qubit device.  For the DC lines that are used to tune the qubit frequencies, we found crosstalk in the 5 qubit device to be typically below about 2\%, which is acceptable because it can be further lowered with careful calibration and digital cancellation \cite{barends}.  In a later 9-qubit chip, the centerline width and gap were respectively reduced from 4\,$\mu$m and 2\,$\mu$m to 3\,$\mu$m and 1.5\,$\mu$m while increasing line separation from 150\,$\mu$m to 200\,$\mu$m, which decreased the crosstalk to about 0.1\% to 1\%.  Microwave crosstalk is still high, around -6 to $-10\,\textrm{dB}$.  In our latest mount we added microwave absorbers inside the mount box.  Even with our various crosstalk models and simulations of box modes, we still need to improve our understanding of this important issue.

\section{Gate Fidelity}

Although $T_1$ and $T_2$ are the key performance metrics in the CSQ program, it is also useful to measure gate fidelity to predict how well qubits should work in a complex quantum algorithm.  Although process tomography has been used in the past to measure fidelity, it has the disadvantages of (1) not reliably accounting for state preparation and measurement (SPAM) errors and (2) requiring increased accuracy of tomography measurements as gate errors are reduced.  Randomized benchmarking (RB) overcomes both of these problems by separating out SPAM errors and amplifying the effect of errors for each repetition of the experiment \cite{barends14}.  This is done by repeating the qubit gate multiple ($n$) times per repetition with a Clifford randomization of the qubit state between each gate; measuring the resultant qubit decay versus $n$ then gives the average gate error.

A current topic of research is whether this simple RB measurement of qubit fidelity is appropriate for predicting the behavior of a complex algorithm.  Initial results by the  UCSB group indicates that indeed this is a good metric.  We have also used RB to tune up the control signals of the qubit to lower gate errors \cite{kelly}.

Measurements of a 5 qubit device \cite{barends14} show consistent fidelity among the different qubits and the complete set of Clifford gates, including the idle (identity) gate, with fidelity scaling as gate time $t_g$.  This shows that the microwave pulses used to generate the gate does not degrade the qubit coherence.  For qubit decay $T_1$, the prediction for gate error is  $\epsilon = t_g/3T_1$; the 1/3 factor comes from averaging over the 6 states arising from the Clifford gates.  The effect of dephasing has a similar dependence.  With gate errors from 0.0006 to 0.001 for 20\,ns gate times, the effective decoherence time is about 8\,$\mu$s, somewhat shorter than predicted for $T_1$, but roughly consistent with decoherence times including dephasing.  (As described previously for $T_2$, a recent experiment has studied gate error versus gate time in detail and explains the qubit fidelities in terms of a new model for dephasing decoherence \cite{omalley}.)

These results for gate fidelity were measured for a multi-qubit device, which includes decoherence effects due to stray coupling of qubits \cite{barends14, omalley}.  Similarly good performance has been observed in our latest 9 qubit device that includes strong coupling for measurement.  Our fidelity numbers are thus indicative of performance for a complex and realistic quantum computing system.

Standard RB uses a Clifford gate set with 90 or 180 degree rotations on the Bloch sphere.  To check for non-regular angles, we have also performed a RB type experiment using a gate set with irrational rotation angles based on Platonic solids \cite{barends14b}.  These results are consistent with values obtained with the Clifford set, showing that fidelity is again not sensitive to the exact choice of gates.

We note that our fidelities were record values when originally published.  Since then, the IBM group has met or slightly exceeded these numbers with single qubit gates, which has come from improving the accuracy of their microwave mixer calibrations, presumably to the degree that was achieved in our control system.

\section{3D versus XMON (IC) transmons}

It is natural to compare the performance of 3D to Xmon (integrated circuit) transmons.  We expect 3D devices to have better raw performance since the size of the devices are larger, lowering the participation ratio and loss from interfaces.  Although transmon coherence times above $100\,\mu\textrm{s}$ has been reported, we caution the reader since at least some of these reports are for transmons with transition frequency much lower than typical values of 5-6 GHz.  For a transition frequency of 2 GHz, the qubit operates 3 times slower, so to compare properly to higher frequency devices the coherence time should be scaled down by a factor of 3.  Thus, these results do not imply better qubit performance or an advance in decoherence.

3D transmon coherence times for $T_1$ is about a factor of 2 longer than for Xmons.  However, when comparing 3D transmons used as multi-qubit devices, the times are comparable or even a bit lower.  We believe such a comparison is better since we are interested in building not single qubit devices but multi-qubit systems.  Comparing $T_2$ for the two types of transmons is more difficult since we have argued previously that this is not a good measure of usable performance.  For this, a suitable metric is probably gate fidelity for both single and coupled qubits.

It is interesting to speculate why 3D performance degrades upon scaling to multi-qubit systems, while not for Xmons.  At UCSB we have used a concept called ``neutrinoization'' to describe what may be behind this.  The idea is that the better coherence of 3D transmons comes from isolating the transmons from the outside world by enclosing them in a box.  However, this box makes it harder to couple to other qubits, so that new decoherence mechanisms may be introduced once qubits are used in more complex cavities with joints. The Xmon has been designed from the start to couple together easily.  Only through good choice of materials and design did we get acceptable coherence, but once this was achieved then scaling is straightforward.

\section{Summary and Outlook}

Building up the qubit technology for a quantum computer is a difficult task.  Making coherent and scalable qubits is not just understanding the basic physics of how qubits work, but also measuring and fixing all the ``dirt physics'' that degrades a particular technology.  For solid-state qubits, this is particularly challenging since the quantum states are spread over a huge number of atoms, allowing them to interact and be disturbed by a large number of defects.  In the presence of non-ideal materials, we still find it astonishing that good coherence is possible when the materials physics is understood and properly designed around.

For superconducing qubits, the two most important defects are two-level states in dielectric insulators and non-equilibrium quasiparticles in superconductors.  By minimizing the use of amorphous dielectrics and choosing qubits that are insensitive to charge, good coherence has been achieved with superconductors.  It should be remembered that qubits that are sensitive to charge, such as the Cooper pair box, have largely been abandoned since they are sensitive to low frequency charge motion, presumably coming from the amorphous materials.

Quasiparticles are expected theoretically to vanish exponentially with temperature, but experimentally we find this is not the case because of non-equilibrium sources of pair-breaking, such as stray infrared light.  For the Cooper pair box, quasiparticles change the charge by $e$ and shift the qubit energy, providing a strong decoherence mechanism.  For qubits not sensitive to charge like the transmon, quasiparticles can still absorb energy to produce qubit decay, but this occurs for only a few percent of tunneling events \cite{martinis09}.  With good infrared filtering the net effect is small, so that devices can be built that are not affected by quasiparticles.

Charge insensitive devices such as the transmon, phase or flux qubits have enabled superconducting qubits to have long coherence times.  Unfortunately, this protection is probably not exhibited for other solid-state systems such as spin qubits or new devices based on Majorana physics.  The interaction of spins in solid state devices is typically moderated by charge in some way, and charge noise is known to be large in these systems and produces rapid dephasing.  We imagine that crystalline dielectrics will reduce charge noise, but it is a challenging materials problem that even the superconducting community has never attempted to solve.  Majorana devices also will probably be sensitive to charge fluctuations.   An even more important consideration for these devices is whether the background population of non-equilibrium quasiparticles will destroy the coherence of this protected topological state.

Although the basic physics of decoherence and materials is now largely understood for superconducting qubits, there remains much research to be done to continue to improve performance.  As we have a good foundation for materials, we are now in a good position to embark upon the next stage of research, scaling up the number of qubits and integrating control wiring and electronics into the chip.  Given the modest size of superconducting qubits and the existence of classical Josephson electronic technology, we are optimistic about this next research frontier: We think it will be as productive as research on decoherence, and scaling will display many of the practical advantages of superconducting qubits.


\begin{thebibliography}{99}
\bibitem{evangeller} Teflon data taken by an undergraduate at UCSB, but not completed and published.  Data is for a resonator made from Nb semirigid coax.  Teflon loss is thought to be very low, but exact value should be considered uncertain.
\bibitem{megrant14} Data taken with MBE Al on a Si substrate, using coplanar resonators with 60\,$\mu$m center trace and 40\,$\mu$m gaps.  At 1 photon power and 50\,mK we achieved an internal intrinsic $Q_i=6\,\times 10^6$.
\end{thebibliography}
\end{document}